\documentstyle[12pt]{article}
\def\be{\begin{eqnarray}}
\def\ee{\end{eqnarray}}
\def\ba{\begin{array}}
\def\ea{\end{array}}
\def\p{\phi}
\def\vp{\varphi}

\def\pa{\partial}

\def\G{{\cal G}}
\def\B{{\cal B}}

\def\X{{\cal X}}

\def\5{^{(5)}}
\begin{document}
\begin{center}
{\LARGE {Solitons via Lie--B\"acklund Transformation for\\
\vskip 0.25cm 5D Low--energy String Theory}}
\end{center}
\vskip 1cm
\begin{center}
{\bf \large {Alfredo Herrera--Aguilar$^\natural$,\footnote{E-mail:
herrera@ifm.umich.mx}}}
\end{center}
\begin{center}
{\bf \large {Joannis E. Paschalis$^\ddag$,\footnote{E-mail:
paschalis@physics.auth.gr}}}
\end{center}
\begin{center}
and
\end{center}
\begin{center}
{\bf \large {Jos\'e Oswald
T\'ellez--V\'azquez$^\natural$,\footnote{E-mail:
oswald@ifm.umich.mx}}}
\end{center}
\vskip 0.1cm
\begin{center}
$^\natural$ Instituto de F\'\i sica y Matem\'aticas, UMSNH\\
Edificio C--3, Ciudad Universitaria, Morelia, Mich. CP 58040
M\'exico
\end{center}
\begin{center}
$^\ddag$ Theoretical Physics Department, Aristotle University of Thessaloniki\\
54124 Thessaloniki, Greece
\end{center}
\vskip 1cm
\begin{abstract}

We apply a non--linear matrix transformation of Lie--B\"acklund type
on a seed soliton configuration in order to obtain a new solitonic
solution in the framework of the 5D low--energy effective field
theory of the bosonic string. The seed solution represents a
stationary axisymmetric two--soliton configuration previously
constructed through the inverse scattering method and consists of a
{\it massless} gravitational field coupled to a non--trivial
chargeless dilaton and to an axion field endowed with charge. We
apply a fully parameterized non--linear matrix transformation of
Ehlers type on this massless solution and get a massive rotating
axisymmetric gravitational soliton coupled to charged axion and
dilaton fields. We discuss on some physical properties of both the
initial and the generated solitons and fully clarify the physical
effect of the non--linear normalized Ehlers transformation on the
seed solution.
\end{abstract}
\newpage
\section{Introduction}

We shall apply a non--linear Lie--B\"acklund transformation in order
to construct solitonic solutions in the framework of the 5D
low--energy effective field theory of the bosonic string. This
theory, when toroidally reduced down to two effective space--time
dimensions (in the presence of $3$ commuting Killing vectors),
allows for the application of the inverse scattering method (ISM),
and hence, for the construction of 5D solitonic field configurations
of gravitational type \cite{yurova}--\cite{haml}. This method was
also modified an implemented for special 4D string models in
\cite{bakyur}. These schemes generalize the ISM originally developed
by Belinski and Zakharov \cite{bz} in the framework of 4D General
Relativity for vacuum stationary axisymmetric gravitational fields
with two Killing vectors. The ISM was further generalized for the
D--dimensional Einstein theory in the presence of $D-2$ Killing
vectors \cite{bv}, and recently applied to obtain the $N$--soliton
static configuration \cite{koikawa} as well as new axisymmetric stationary
solutions \cite{mishima-iguchi} in the framework of 5D Einstein
theory, Einstein--Maxwell \cite{ISM-EM} and the
Einstein--Maxwell with dilaton systems \cite{azuma-koikawaEMD}.

In the framework of D--dimensional low--energy effective string
theories the problem becomes more complicated since the chiral
representations of these field theories have dimensions greater than
two and must satisfy non--trivial group conditions. The chiral
matrix which parameterizes the low--energy effective field theory of
the heterotic string when reduced to two space--time dimensions
possesses the $SO(d+1,d+n+1)/[SO(d+1)\times SO(d+n+1)]$ symmetry
group \cite{ms}, where $d$ is the number of compactified dimensions
and $n$ is the number of Abelian vector fields of the theory. This
chiral representation involves matrices of dimension $(2d+n+2)$
which must satisfy orthogonal group conditions. As a consequence,
the original scheme of Belinski and Zakharov must be suitably
modified and generalized \cite{yurova}. However, it is not a trivial
task to overcome this problem since the group symmetry condition of
the chiral matrices and their dimensionality strongly restrict the
solitons we can construct, eventually leading to even trivial
solutions.

The physical interpretation of the seed soliton we shall use is
quite interesting since it describes a massless gravitational
object\footnote{More precisely, an object without mass term in the
asymptotical expansion of the $g_{tt}$--component.} coupled to
non--trivial dilaton and Kalb--Ramond fields, which possesses,
however, a term which describes angular momentum multiplied by a NUT
parameter. Another novel feature of the solitonic seed solution
appears when looking towards its rotating black hole/string
interpretation; in order to achieve this aim, we must impose the
asymptotic flatness condition and set to zero the NUT parameter.
However, when one drops the NUT charge, the field configuration
becomes static. Thus, rotating solitons necessarily possess a NUT
parameter, and, asymptotically flat configurations are necessarily
static. In \cite{haml}, a seed solution of this kind was endowed
with gravitational mass and dilaton charge with the aid of a {\it
simplified} non--linear transformation of Lie--B\"acklund type
\cite{stephani}, the so--called normalized Ehlers transformation
(NET), curing in this way its first strange property. However, the
second anomaly remained present in the constructed field
configuration since the NET did not affect the physical
interpretation of the rotating functions of the metric.

In this paper we apply a fully parameterized NET \cite{hk5} on the
massless seed solution constructed in \cite{haml} in order to get a
new solitonic object. In this way we can clarify the physical effect
of the NET and make an interpretation of the obtained field
configuration. These stringy exact solutions are of interest because
they can represent new objects that can properly be interpreted in
the framework of string theory since in 5D General Relativity there
are no uniqueness theorems and different objects, black holes and
rings for instance, can be constructed and identified by the same
set of parameters.

\section{Low--energy Effective Action and MEP}
In this Section we shall briefly review the field content the 5D
low--energy effective field theory of the bosonic string as well as
its stationary matrix Ernst potential (MEP) formulation. Then we
point out the nonlinear NET symmetry of this stationary theory
written in the language of the MEP. Let us start by considering the
following action \be S\5\!=\!\int\!d^5\!x\!\mid\!
G\5\!\mid^{\frac{1}{2}}\!e^{-\p\5}\left(R\5\!+\!
\p\5_{;M}\,\p^{(5);M}
\!-\!\frac{1}{12}\!H\5_{MNP}\,H^{(5)MNP}\right), \ee where
$G\5_{MN}$ is the metric, $H\5_{MNP}=\pa_MB\5_{NP} + \mbox{{\rm
cycl. perms. of} M,\,N,\,P;}$ $B\5_{MN}$ is the anti--symmetric
Kalb--Ramond tensor field, $\p\5$ is the dilaton and
$M,N,P=1,2,...,5$. The stationary theory which results after the
Kaluza--Klein compactification of this model on $T^2$ possesses the
$SO(3,3)/[SO(3)\times SO(3)]$ symmetry group and describes an
effective three--dimensional gravity parameterized by the metric
tensor \be g_{\mu\nu}=e^{-2\p}\left[G_{\mu,\nu}\5-
G_{p+3,\mu}\5G_{q+3,\nu}\5\left(G^{-1}\right)_{pq}\right] \ee
coupled to the following fields: the three--dimensional dilaton
field $\p$, and the moduli $2\times2$--matrices $G\equiv G_{pq}$ and
$B\equiv B_{pq}$ \be \p=\p\5-\frac{1}{2}{\rm ln}\mid{\rm det}G\mid,
\qquad G_{pq}=G_{p+3,q+3}\5, \qquad B_{pq}=B_{p+3,q+3}\5,
\label{GBp} \ee where $p,q=1,2$ label the time and the fifth
coordinate, respectively; the vector fields represented by the
$2\times 3$--matrices \be
(A_1)_{\mu}^p=\frac{1}{2}\left(G^{-1}\right)_{pq}G_{q+3,\mu}\5,
\qquad (A_2)_{\mu}^{p+2}=\frac{1}{2}B_{p+3,\mu}\5-B_{pq}A_{\mu}^q,
\ee where $\mu,\nu=1,2,3$; and the antisymmetric tensor field
$B_{\mu\nu}$ (which we set to zero since it does not possess
dynamical properties in three dimensions) \be
B_{\mu\nu}=B_{\mu\nu}\5-4B_{pq}A_{\mu}^pA_{\nu}^q-
2\left(A_{\mu}^pA_{\nu}^{p+2}-A_{\nu}^pA_{\mu}^{p+2}\right). \ee

In three spatial dimensions, the vector fields can be dualized
on--shell through the pseudoscalar fields $u$ and $v$ in the
following form
\begin{eqnarray}
\nabla\times\overrightarrow{A_1}&=&\frac{1}{2}e^{2\p}G^{-1}
\left(\nabla u+B\nabla v\right),
\nonumber                          \\
\nabla\times\overrightarrow{A_2}&=&\frac{1}{2}e^{2\p}G\nabla v-
B\nabla\times\overrightarrow{A_1},\label{dual}
\end{eqnarray}
where all vector and differential operations are performed with
respect to the three--dimensional metric $g_{\mu\nu}$.

Thus, the effective stationary theory describes three--dimensional
gravity $g_{\mu\nu}$ coupled to the scalar fields $G$, $B$, $\p$ and
to the pseudoscalar ones $u$, $v$.

In \cite{hk3} it was shown that all these matter fields can be
arranged in the following $3\times3$--matrix Ernst potential \be
\X=\left( \ba{cc} -e^{-2\p}+v^TXv&v^TX-u^T \cr Xv+u&X \ea \right),
\label{X}\ee where $X$ is a $2\times2$--matrix potential $X=G+B$.
The potential $\X$ was called matrix Ernst potential due to its
close analogy to the complex Ernst potential which enters in the
stationary formulation of the vacuum Einstein theory \cite{e}, which
was further generalized for low--energy string theory \cite{ggk}.
The physical meaning of its components is the following: the
relevant information concerning the gravitational field is encoded
in the matrix potential $X$ through the moduli matrix $G$, whereas
its rotational character is parameterized by the dualized variable
$u$; $X$ also contains the antisymmetric Kalb--Ramond tensor field
$B$, whereas its multidimensional components are dualized through
the pseudoscalar field $v$, and the 3D dilaton field is represented
by $\phi$.

In terms of the MEP the effective stationary theory adopts the
following form \be ^3S\!= \!\int\!d^3x\!\mid
g\mid^{\frac{1}{2}}\!\left[\!-\!R\!+ \!{\rm Tr}\left(\frac{1}{4}\
\nabla\X\G^{-1}\nabla\X^T\G^{-1}\right)\right],\label{acX}\ee where
$\G=\left(\X+\X^T\right)/2$ and $\B=\left(\X-\X^T\right)/2$. The MEP
$\X$ also is the sum of its symmetric and antisymmetric parts:
$\X=\G+\B$ \, where \be \G=\left( \ba{cc} -e^{-2\p}+v^TGv&v^TG \cr
Gv&G \ea \right), \quad \B=\left( \ba{cc} 0&v^TB-u^T \cr Bv+u&B \ea
\right). \label{GyB}\ee

In \cite{hk5} it was performed a full classification of the
symmetries of the stationary action (\ref{acX}) in the language of
the MEP according to their charging properties. Among them one
encounters a {\it hidden} symmetry that was called normalized Ehlers
transformation. This non--linear symmetry acts in a non--trivial way
on the properties of space--time when applying it on a seed solution
of the theory, both in the framework of General Relativity and
string theory. For instance, by applying the Ehlers transformation
on both vacuum Schwarzschild and Kerr solutions, it generates the
so--called NUT charge, changing the topological properties of the
space--time \cite{kramer}. The NET is a matrix generalization of the
Lie--B\"acklund symmetry found by Ehlers in the vacuum Einstein
theory.

The matrix NET transformation reads  \be
\X\rightarrow\left(1+\Sigma\Lambda\right)
\left(1+\X_0\Lambda\right)^{-1} \X_0\left(1-\Lambda\Sigma\right)+
\Sigma\Lambda\Sigma,\label{net} \ee where $\Sigma=diag(-1,-1;1)$ and
$\Lambda$ is an arbitrary antisymmetric constant $3\times 3$--matrix
$\Lambda=-\Lambda^T$. Since this a symmetry of the stationary action
(\ref{acX}), it can be used to generate new solutions starting from
simple initial solutions. Thus, by applying the NET on a seed
solution, we obtain a new stationary solution endowed with three
more parameters introduced through the antisymmetric matrix
$\Lambda$. It is interesting to see what is the physical effect of
the NET, and if the introduced parameters can eliminate some of the
strange features of the solitonic seed solution we shall consider in
the next Section.
\section{Solution--generating technique}
In this Section we shall apply the non--linear NET on a
solitonic seed configuration constructed through the ISM in order to
analyze the physical effect of the NET and to compare the physical
properties of both the initial and the generated solutions.

We first present the explicit expression for the two--soliton seed
solution constructed through the ISM in \cite{haml}. The 5D metric
reads \be
ds_5^2=G_{pq}(dx^p-\omega_{\vp}^pd\vp)(dx^q-\omega_{\vp}^qd\vp)+e^{2\p}ds_3^2,
\label{interval}\ee where the components of the moduli matrix
$G_{pq}$ are \be G_{11}=-\frac{r^2+b^2-(c-a\cos\theta)^2}{\delta^2},
\quad\quad G_{12}=\frac{2cr}{\delta^2}, \quad\quad
G_{22}=\frac{r^2+b^2-(c+a\cos\theta)^2}{\delta^2}, \label{Gpq}\ee
the metric functions $\omega_{\vp}^q$ are defined by the relations
\be
\omega_{\vp}^1=\frac{-2a\sqrt{bc}\,r\sin^2\theta}{\Delta+a^2\sin^2\theta},
\quad\quad \omega_{\vp}^2=2\sqrt{bc}\left[\cos\theta+
\frac{a\left(b+c-a\cos\theta\right)\sin^2\theta}{\Delta+a^2\sin^2\theta}\right],
\label{oms}\ee the three--dimensional dilaton field has the form \be
e^{2\p}=1-\frac{4bc}{\Delta+a^2\sin^2\theta}, \label{3dilaton}\ee
the spatial line element is
\begin{eqnarray}
ds_3^2=\left(\Delta+a^2\sin^2\theta\right)\left[\frac{dr^2}{\Delta}+
d\theta^2\right]+\Delta\sin^2\theta d\varphi^2.
\end{eqnarray}
The components of the antisymmetric Kalb--Ramond tensor field are
\be B_{12}\!=\!\frac{2br}{\delta^2},\quad
B_{4,\vp}\5\!=\!\frac{2\sqrt{bc}\,r(2b\cos\!\theta\!+\!a\sin^2\!\theta)}{\delta^2},\quad
B_{5,\vp}\5\!=\!2\sqrt{bc}\!\left[\!\cos\!\theta\!-
\!\frac{a(c\!-\!b\!+\!a\cos\!\theta)\!\sin^2\!\theta}{\delta^2}\!\right]\!\!
\label{B12}\ee and the 5D dilaton field reads \be
e^{\p\5}=\frac{r^2+(b-c)^2-a^2\cos^2\theta}{\delta^2}.
\label{5dilaton}\ee Here we have introduced the following notation:
$\sigma^2=a^2-(b+c)^2$, $\Delta=r^2-\sigma^2$ and
$\delta^2=r^2+c^2-(b-a\cos\theta)^2$, where $a$, $b$ and $c$ are
constants.

This field configuration represents a stationary axially symmetric
{\it massless} gravitational field coupled to a non--trivial
chargeless dilaton field and to an axion field endowed with charge.
Despite the massless character of the gravitational field, it
possesses angular momentum corresponding to the rotation in the $\vp
t$--plane. Namely, by studying the asymptotical behaviour of the
metric components $\omega_{\vp}^q$ one sees that $\omega_{\vp}^1$
defines the following angular momentum \be
\omega_{\vp}^1\mid_{r\rightarrow\infty}
\,\,\,\sim\frac{-2a\sqrt{bc}\sin^2\theta}{r}, \label{om1}\ee whereas
$\omega_{\vp}^2$ is not asymptotically flat and defines an effective
NUT--like parameter: \be \omega_{\vp}^2\mid_{r\rightarrow\infty}
\,\,\,\sim 2\sqrt{bc}\cos\theta. \label{om2}\ee

As pointed out above, as soon as one requires asymptotic flatness,
the field configurations become static, whereas the NUT parameter is
always present in these stationary solitons. Thus, our family of
solutions does not contain a spherically symmetric subclass. It is
interesting to note that a comparison to previously found
configurations \cite{othersols} shows that this family of solutions
cannot be obtained by setting to zero the mass or other parameters.

\subsection{New solitonic configuration via NET}

In \cite{haml} it was proved that the generalization of the ISM
proposed by Yurova \cite{yurova} cannot provide gravitational
solitons with mass term. Moreover, in that work, a simplified
version of the NET was implemented to cure this strange feature of
these ISM solitonic configurations. Here we shall apply the fully
parameterized Lie--B\"acklund transformation NET (\ref{net}) on a
seed solution which corresponds to the massless gravitational
soliton (\ref{interval})--(\ref{5dilaton}).

Thus, the seed MEP $\X_0=\G_0+\B_0$ corresponding to this field
configuration reads \be \X_0=\delta^{-2}\left( \ba{ccc} -\delta^2
&0& 0\cr
4\sqrt{bc}(b+c-a\cos\theta)&-\left[r^2+b^2-(c-a\cos\theta)^2\right]
&2(b+c)r \cr 4\sqrt{bc}r &2(c-b)r &r^2+b^2-(c+a\cos\theta)^2 \ea
\right)\!\!, \label{X0}\ee where $\G_0$ and $\B_0$ are seed matrix
potentials that correspond to (\ref{GyB}).

Let us make use of the fully parameterized antisymmetric matrix
$\Lambda$: \be \Lambda=\left( \ba{ccc} 0&k_1&k_2 \cr -k_1&0&k_3\cr
-k_2&-k_3 & 0\ea \right), \label{l}\ee where $k_1$, $k_2$ and $k_3$
are arbitrary constants.

After applying the NET on $\X_0$ with this matrix $\Lambda$ we
obtain a new solution of the theory endowed with three more
parameters. Thus, after tedious and lengthy intermediate
calculations, the components of the transformed matrix $G_{pq}$
adopt the form
\begin{eqnarray}
G_{11}&=&-\frac{K_1\left(r^2-a^2\cos^2\theta\right)+4\left(K_2+K_3\right)r+
2\left(K_5+2K_7\right)a\cos\theta+(b+c)K_8}{K_1\left(r^2-a^2\cos^2\theta\right)+
4\left(K_2-K_4\right)r+2\left(K_6-2K_7\right)a\cos\theta-(b+c)K_8},
\nonumber\\
G_{12}&=&\frac{2\left[\left(K_5+2K_7\right)r+2\left(K_2+K_3\right)a\cos\theta\right]}
{K_1\left(r^2-a^2\cos^2\theta\right)+
4\left(K_2-K_4\right)r+2\left(K_6-2K_7\right)a\cos\theta-(b+c)K_8},\\
G_{22}&=&\frac{K_1\left(r^2-a^2\cos^2\theta\right)-4\left(K_2+K_3\right)r-
2\left(K_5+2K_7\right)a\cos\theta+(b+c)K_8}
{K_1\left(r^2-a^2\cos^2\theta\right)+
4\left(K_2-K_4\right)r+2\left(K_6-2K_7\right)a\cos\theta-(b+c)K_8},
\nonumber\label{G'pq}\end{eqnarray} where
$K_1\!=\!\left(1\!+\!k_1^2\!-\!k_2^2\!-\!k_3^2\right)$,
$K_2\!=\!(k_2+k_1k_3)\sqrt{bc}$, $K_3\!=\!ck_3+bk_1k_2$,
$K_4\!=\!bk_3+ck_1k_2$, $K_5=(k_1^2+k_2^2)b+(1+k_3^2)c$,
$K_6=(1+k_3^2)b+(k_1^2+k_2^2)c$, $K_7=(k_1+k_2k_3)\sqrt{bc}$, and
$K_8=\left(b\!-\!c\right)\left(1\!-\!k_1^2\!+\!k_2^2\!-\!k_3^2\right)
-4\left(k_1-k_2k_3\right)\sqrt{bc}\ $; the metric functions
$\omega_{\vp}^q$ read
\begin{eqnarray} \omega_{\vp}^1&=&\frac{2K_{10}}{K_1}\cos\theta-
\frac{2a\left[K_9r+K_{10}\,a\cos\theta-K_{11}\right]\sin^2\theta}
{K_1\left(\Delta+a^2\sin^2\theta\right)}, \nonumber\\
\omega_{\vp}^2&=&\frac{2K_9}{K_1}\cos\theta\!-\!\frac{2a\left[K_{10}\,r\!+\!K_9\,a\cos\theta\!-\!K_{12}\right]\sin^2\theta}
{K_1\left(\Delta+a^2\sin^2\theta\right)}, \label{oms'}\end{eqnarray}
where
$K_9\!=\!(k_1\!-\!k_2k_3)\!(b\!-\!c)\!+\!\left(1\!-\!k_1^2\!+\!k_2^2\!-\!k_3^2\right)\!\sqrt{bc}$,
$K_{10}\!=\!(k_2\!-\!k_1k_3)\!(b\!+\!c)$,
$K_{11}\!=\!(b\!+\!c)\!\left[2(k_3\!-\!k_1k_2)\!\sqrt{bc}\!+\!(k_2+k_1k_3)\!(b\!-\!c)\right]$,
$K_{12}\!=\!(b\!+\!c)\left[\left(1\!-\!k_1^2\!-\!k_2^2\!+\!k_3^2\right)\sqrt{bc}\!+\!(k_1\!+\!k_2k_3)(b\!-\!c)\right]$;
the 3D dilaton field is given by the expression \be
e^{2\p}=1+\frac{K_8^2-(b+c)^2K_1^2}{K_1^2\left(\Delta+a^2\sin^2\theta\right)},
\label{3dilatonnet}\ee which is still chargeless; the components of
the antisymmetric Kalb--Ramond field are
\begin{eqnarray}
B_{12}&=&\frac{2\left[\left(K_6-2K_7\right)r+2\left(K_2-K_4\right)a\cos\theta\right]}
{K_1\left(r^2-a^2\cos^2\theta\right)+
4\left(K_2-K_4\right)r+2\left(K_6-2K_7\right)a\cos\theta-(b+c)K_8},
\nonumber\\
\end{eqnarray}
and the 5D dilaton field is given by the following relation \be
e^{\p\5}=\frac{K_1^2\left(r^2-a^2\cos^2\theta\right)+K_8^2}
{K_1\left[K_1\left(r^2-a^2\cos^2\theta\right)+4\left(K_2-K_4\right)r+2\left(K_6-2K_7\right)a\cos\theta-(b+c)K_8\right]}.
\label{5dilatonnet}\ee As a checking exercise, it is straightforward
to see that when the $k_i=0$ ($i=1,2,3$), one recovers the seed
solitonic configuration (\ref{interval})--(\ref{5dilaton}), and when
only $k_1$ and $k_2$ vanish, one gets the solution generated in
\cite{haml}.

By studying the asymptotic properties of our generated field
configuration one can observe that the gravitational components
$G_{pq}$ have acquired mass terms \be
G_{11}\mid_{r\rightarrow\infty}\sim -1+\frac{2m_{11}}{r}, \quad\quad
G_{12}\mid_{r\rightarrow\infty}\sim -\frac{m_{12}}{r}, \quad\quad
G_{22}\mid_{r\rightarrow\infty}\sim 1-\frac{2m_{22}}{r},
\label{G'pqinfty}\ee where we have introduced the following
effective masses \be
m_{11}=\frac{2(K_3+K_4)}{\left(k_2^2\!+\!k_3^2\!-\!1\!-\!k_1^2\right)},
\quad\quad
m_{12}=\frac{2\left(2K_7+K_5\right)}{\left(k_2^2\!+\!k_3^2\!-\!1\!-\!k_1^2\right)},
\quad\quad
m_{22}=\frac{2\left(K_4-2K_2-K_3\right)}{\left(k_2^2\!+\!k_3^2\!-\!1\!-\!k_1^2\right)},\ee
whereas the metric functions $\omega_{\vp}^q$ adopt the following
asymptotic form
\begin{eqnarray} \omega_{\vp}^1\mid_{r\rightarrow\infty}&\sim&\frac{2K_{10}}{K_1}\cos\theta-
\frac{2aK_9\sin^2\theta}{K_1r},\\
\omega_{\vp}^2\mid_{r\rightarrow\infty}&\sim&
\frac{2K_9}{K_1}\cos\theta\!-\!\frac{2aK_{10}\,\sin^2\theta}{K_1r},
\nonumber \label{omstransf}\end{eqnarray} making manifest the
appearance of an angular momentum term corresponding to the rotation
in the $\vp x^5$--plane and an effective NUT charge in the $\vp
t$--plane. The transformed component $B_{12}$ of the antisymmetric
tensor field and the 5D dilaton asymptotically behave as  \be
B_{12}\mid_{r\rightarrow\infty}\sim\frac{b_{12}}{r}, \quad\quad
e^{\p\5}\mid_{r\rightarrow\infty}\sim 1+\frac{D}{r}, \ee where we
have introduced the following charges \be
b_{12}=\frac{2\left(2K_7-K_6\right)}{\left(k_2^2\!+\!k_3^2\!-\!1\!-\!k_1^2\right)},
\quad\quad
D=\frac{4\left(K_2-K_4\right)}{\left(k_2^2\!+\!k_3^2\!-\!1\!-\!k_1^2\right)}.
\label{B'fi'infty}\ee

The explicit expressions for the $B_{3+p,\vp}\5$ components of the
Kalb--Ramond field are quite involved and we shall omit them,
however, we point out that their asymptotic behaviour is very
similar to that of the metric functions $\omega_{\vp}^q$, underlying
the torsion nature of this tensor field.


Thus, the solitonic solution we have constructed represents a 5D
stationary axisymmetric massive gravitational field configuration
coupled to charged dilaton and axion fields. This metric possesses
two angular momenta as well as two NUT effective parameters defined
in the $\vp t$-- and $\vp x^5$-- planes. By imposing asymptotic
flatness we must set to zero $K_9$ and $K_{10}$, however, in this
limit the field configuration becomes static. If just $K_{10}=0$
($K_9=0$), we obtain a metric with angular momentum corresponding to
rotation in the $\vp t$-- ($\ \vp x^5$--) plane, but with NUT
parameter in the $\vp x^5$-- ($\ \vp t$--) plane, as the metric of
the initial solution.

\section{Concluding remarks}

In this paper we have constructed a solitonic gravitational field
configuration to the 5D low--energy bosonic sector of string theory
by applying a non--linear matrix transformation of Lie--B\"acklund
type. As a seed solution we used a rotating massless gravitational
field configuration coupled to non--trivial dilaton and axion fields
which has been obtained previously by means of the ISM. This
massless field configuration possesses some strange physical
properties. We further endowed this gravitational seed configuration
with mass and charge terms, and cured some of its strange physical
peculiarities by performing a fully parameterized NET on it.

Thus, the physical meaning of the NET consists of providing masses
and charges for the 5D fields as well as generating an angular
momentum term and a NUT parameter in the $\vp x^5$-- and the $\vp
t$--planes, respectively. One more time we see that when we impose
the asymptotically flatness condition, we obtain a static field
configuration since by setting to zero the effective NUT parameters
$K_9$ and $K_{10}$, we force the metric functions (\ref{oms'}) to
vanish. Thus, if our soliton represents a rotating field
configuration, it necessarily possesses NUT parameters and if it
constitutes an asymptotically flat solution, it is necessarily
static, a feature which is not shared by rotating configurations in
General Relativity.

Thus, by comparing the seed and the generated solitonic
configurations we see that they share the strange property of being
stationary, but with NUT charges, or asymptotically flat, but
static. This feature deserves a deeper investigation since both the
ISM and the implementation of Lie--B\"acklund transformations could
lead to the construction of new charged black hole (brane) solutions
in low--energy string theories formulated in $D>4$ dimensions as the
non--asymptotically flat black hole configurations constructed in
\cite{yazad}.

\section*{Acknowledgments}

One of the authors (AHA) would like to express his deepest and
sincere gratitude to the organizers and all the participants of the
International Conference on Geometry, Dynamics and Integrable
Systems 2008 for providing a warm and friendly atmosphere during the
meeting in Beograd, Serbia. This research was supported by grants
CIC-UMSNH-4.16 and CONACYT-60060.


\begin{thebibliography}{100}

\bibitem {yurova}
M.V. Yurova, Phys. Rev. D {\bf 64} 024022 (2001).

\bibitem {haml}
A. Herrera--Aguilar and R.R. Mora--Luna, Phys. Rev. D {\bf 69}
105002 (2004).

\bibitem {bakyur} I. Bakas, Nucl. Phys. {\bf B428} 374
(1994); Phys. Rev. D {\bf 54} 6424 (1996); H.J. de Vega and N.
S\'anchez, Phys. Rev. {\bf D47} 3394 (1993); F. Combes, H.J. de
Vega, A.V. Mikhailov and N. S\'anchez, Phys. Rev. {\bf D50} 2754
(1994); A. Eris, A. Karasu and M. Gurses, J. Math. Phys. {\bf 25}
1489 (1984); M.V. Yurova, Gen. Rel. Grav. {\bf 32} 2219 (2000);
Phys. Rev. {\bf D65} 024024 (2001); O.V. Kechkin, Class. Quantum
Grav. {\bf 20} 2157 (2003); {\bf 20} L225 (2003); G.A. Alekseev and
M.V. Yurova, ``Integrable structure of the low--energy string
gravity equations in $D=4$ space-times with two commuting
isometries", hep--th/0401077.

\bibitem {bz}
V.A. Belinski and V.E. Zakharov, Sov. Phys. JETP {\bf 48} 985
(1978); {\bf 50} 1 (1979).

\bibitem{bv}
V. Belinski and E. Verdaguer, {\it Gravitational Solitons},
Cambridge University Press (Cambridge, 2001).

\bibitem {koikawa} T. Koikawa, Prog. Theor. Phys. {\bf 114} 793 (2005).

\bibitem {mishima-iguchi} T. Mishima and H. Iguchi, Phys. Rev. {\bf D73} 044030 (2006);
Phys. Rev. {\bf D73} 121501 (2006); Phys. Rev. {\bf D74} 024029  (2006);
T. Azuma and T. Koikawa, Prog. Theor. Phys. {\bf 116} 319 (2006); S. Tomizawa, Y. Morisawa
and Y. Yasui Phys. Rev. {\bf D73} 064009 (2006); S. Tomizawa and M. Nozawa,
Phys. Rev. {\bf D73} 124034 (2006).

\bibitem {ISM-EM} S.S. Yazadjiev, Phys. Rev. {\bf D73} 104007 (2006);
T. Azuma and T. Koikawa, Prog. Theor. Phys. {\bf 118} 35 (2007).

\bibitem {azuma-koikawaEMD} T. Azuma and T. Koikawa, Progr. Theoret. Phys. {\bf 121} 627 (2009).

\bibitem {ms}
N. Marcus and J.H. Schwarz, Nucl. Phys. {\bf B228} 145 (1983); J.
Maharana and J.H. Schwarz, Nucl. Phys. {\bf B390} 3 (1993); A.
Sen, Nucl. Phys. {\bf B434} 179 (1995).

\bibitem{stephani}
H. Stephani, {\it Differential Equations and their Solutions Using
Symmetries}, Cambridge University Press (Cambridge, 1987).

\bibitem {hk5}
A. Herrera--Aguilar and O. Kechkin,  Phys. Rev. {\bf D59} 124006
(1999).

\bibitem {hk3}
A. Herrera--Aguilar and O. Kechkin, Int. J. Mod. Phys. {\bf A13} 393
(1998); {\bf A14} 1345 (1999).

\bibitem {e}
F.J. Ernst, Phys. Rev. {\bf 168} 1415 (1968); {\bf 167} 1175 (1968);
J. Math. Phys. {\bf 12} 2395 (1971).

\bibitem {ggk}
D.V. Gal'tsov and O. Kechkin,  Phys. Rev. {\bf D50} 7394 (1994); D.
Gal'tsov, A.A. Garc\'\i a, and O. Kechkin, J. Math. Phys. {\bf 36}
5023 (1995).

\bibitem {kramer}
D. Kramer, H. Stephani, M. McCallum and E. Herlt, {\it Exact
Solutions of the Einstein's Field Equations}, Deutcher Verlag der
Wissenschaften (Berlin, 1980).

\bibitem {othersols}
G.T. Horowitz and A. Sen, Phys. Rev. {\bf D54} 808 (1996); M. Cvetic
and D. Youm, Nucl. Phys. {\bf B476} 118 (1996); A.A. Tseytlin, Mod.
Phys. Lett. A {\bf 11} 689 (1996); R. Kallosh, A. Rajaraman and W.K.
Wong, Phys. Rev. {\bf D55} 3246 (1997); J.C. Breckenridge, R.C.
Myers, A.W. Peet and C. Vafa, Phys. Lett. {\bf B391} 93 (1997); J.C.
Breckenridge, D.A. Lowe, R.C. Myers, A.W. Peet, A. Strominger and C.
Vafa, Phys. Lett. {\bf B381} 423 (1996); A. Herrera--Aguilar and O.
Kechkin, Mod. Phys. Lett. {\bf A13} 1629 (1998); C.A.R. Herdeiro,
Nucl. Phys. {\bf B582} 363 (2000); T. Matos, D. N\'u\~nez, G.
Est\'evez and M. R\'\i os, Gen. Rel. Grav. {\bf 32} 1499 (2000); R.
Emparan and H.S. Reall, Phys. Rev. Lett. {\bf 88} 4877 (2002); H.
Elvang, Phys. Rev. {\bf D68} 124016 (2003); S.S. Yazadjiev, Phys.
Rev. {\bf D72} 104014 (2005); Phys. Rev. {\bf D73} 064008 (2006);
Phys. Rev. {\bf D77} 127501 (2008); R. Emparan and H.S. Reall,
Living Rev. Rel. {\bf 11} 6 (2008).

\bibitem {yazad} S.S. Yazadjiev, Class. Quant. Grav. {\bf 22} 3875 (2005);
Phys. Rev. {\bf D73} 124032 (2006).


\end{thebibliography}
\end{document}